# Giant anomalous Hall conductivity in the itinerant ferromagnet LaCrSb$_3$ and the effect of *f*-electrons

*Nitesh Kumar, Neetu Lamba, Jacob Gayles, Congcong Le, Praveen Vir, Satya N. Guin, Yan Sun, Claudia Felser, Chandra Shekhar*

Max Planck Institute for Chemical Physics of Solids, 01187 Dresden, Germany

Email: shekhar@cpfs.mpg.de

**Keywords**: anomalous Hall effect, itinerant ferromagnet, electrical transport, first principles calculations

**Abstract**
Itinerant ferromagnets constitute an important class of materials wherein spin-polarization can affect the electric transport properties in nontrivial ways. One such phenomenon is anomalous Hall effect which depends on the details of the band structure such as the amount of band crossings in the valence band of the ferromagnet. Here, we have found extraordinary anomalous Hall effect in an itinerant ferromagnetic metal LaCrSb$_3$. The rather two-dimensional nature of the magnetic subunit imparts large anisotropic anomalous Hall conductivity of 1250 $\Omega^{-1}$cm$^{-1}$ at 2K. Our investigations suggest that a strong Berry curvature by abundant momentum-space crossings and narrow energy-gap openings are the primary sources of the anomalous Hall conductivity. An important observation is the existence of quasi-dispersionless bands in LaCrSb$_3$ which is now known to increase the anomalous Hall conductivity. After introducing *f*-electrons, anomalous Hall conductivity experiences more than two-fold increase and reaches 2900 $\Omega^{-1}$cm$^{-1}$ in NdCrSb$_3$.

**Introduction**

Nontrivial band topology features a unique electronic structure that describes the origin of the quantum Hall effect, which exists with many variants. In the Hall effect, a mutually perpendicular magnetic field and electric current applied in materials causes a voltage perpendicular to them, i.e., the Hall voltage. Similarly, a comparatively large spontaneous Hall effect, termed the anomalous Hall effect (AHE), is known to exist in magnetic materials in which the Bloch wave function of electrons is asymmetric in momentum space. In this scenario, electrons acquire an additional group velocity in the presence of a driving perturbation, such as an external electric field. This anomalous velocity is perpendicular to the applied electric field, giving rise to an additional value to the Hall effect, i.e., AHE.[1] In addition, the group velocity is drastically enhanced by virtue of the Berry phase of nontrivial bands, which provides a strong fictitious field.

A nontrivial band topology arises when band inversion occurs, i.e., the conduction band is beneath the valance band with respect to their natural order in the vicinity of the Fermi level $E_F$ (**Figure 1**a, left). Such inversion can be of several possible combinations among the $s-$, $p-$, $d-$, and $f-$ bands. Figure 1a (right) shows a schematic of various types of band mixing; a resulting band gap arises after considering spin orbit coupling (SOC) regardless of their type. In such cases, the wave function of each band twists in momentum space inducing a non-zero Berry phase. A linear response of conductivity [2, 3] from the Berry phase [4] for a 3D system is expressed in the Kubo formula:

$$\sigma_{ij} = \frac{e^2}{\hbar} \sum_n \int \frac{d^3k}{(2\pi)^3} \Omega_{ij}^n f(\varepsilon_k)$$

$\Omega_{ij}^n$ is the Berry curvature, which crucially depends on the entanglement of bands. A small convergence of bands is caused by a large contribution from mixed occupied states, whereas its counterpart, i.e., unoccupied states, contribute negligibly. Therefore, AHE is rather large when the SOC-induced gap is small in the vicinity of $E_F$. The material selection and desired band gap depend strongly upon the hybridization strength (by the lattice constant) and the magnitude of SOC (by the atomic charge). Interestingly, the dispersion of bands is a crucial factor and further depends on orbital hybridizations. For example, Figure 1b shows the calculated electronic band structure of LaCrSb$_3$, wherein several bands are relatively dispersionless, quite close to $E_F$ along $Y$-$\Gamma$. Such bands are highly sensitive to perturbation and account for various intriguing phenomena such as nontrivial topology [5-7], high-temperature fractional Hall effect [6, 8-11], unconventional superconductivity [12, 13], and unconventional magnetism.[7, 14, 15]

In our present selection of compound i.e. LaCrSb$_3$ the quasi-dispersionless bands facilitate larger mixing of occupied and unoccupied bands close to $E_F$, that induces a large volume of Berry curvatures (BCs). Such BC associated with nontrivial bands as a source of AHE has recently been recognized in various compounds, for example, chiral antiferromagnets Mn$_3$Sn and Mn$_3$Ge [16, 17], ferromagnetic massive Dirac metal Fe$_3$Sn$_2$ [18], and ferromagnetic nodal line compounds Co$_2$MnGa [19, 20], Co$_3$Sn$_2$S$_2$ [21] and Fe$_3$GeTe$_2$.[22]

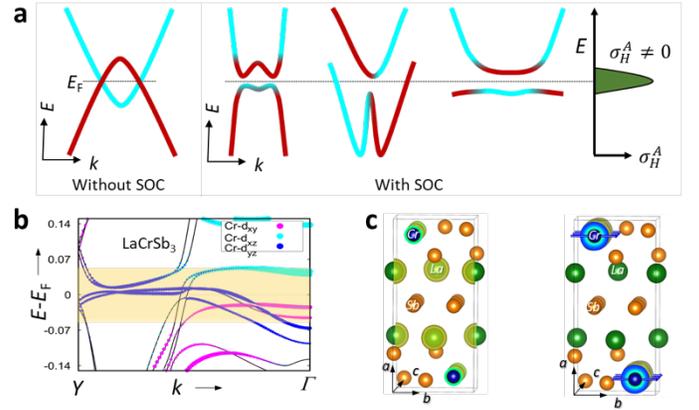

*Figure 1. Nontrivial origins of anomalous Hall conductivity, $\sigma_H^A$ band structure Berry curvature (BC), and charge and spin densities. a) Schematic of different types of overlapping bands that result in zero (left) and non-zero (right) values of intrinsic $\sigma_H^A$, depending on spin orbit coupling (SOC). b) Orbital dependent electronic band structure of LaCrSb$_3$ along Y-Γ with SOC, depicting (shaded region) various quasi-dispersionless bands (very small change of energy with momentum) quite close to the Fermi energy, $E_F$. c) Charge density (left) and spin density (right) localization in LaCrSb$_3$.*

## Results

LaCrSb$_3$ possesses an orthorhombic centrosymmetric crystal structure and belongs to the *Pbcm* space group (no. 57). The lattice parameters $a$, $b$, and $c$ for LaCrSb$_3$ are 13.18, 6.16, and 6.07 Å, respectively and they decrease smoothly on replacing rare earth metals from La to Lu except for Yb.[23] In the crystal structure as shown in Figure S1a, the atoms are arranged in special manners: The Sb square nets form perpendicular to the [100]; the edge- and face-sharing of CrSb$_6$ octahedra arrange along the $b-$ and $c-$axes, respectively. CrSb$_2$–magnetic layers in the $b$–$c$ plane gives rise to a 2D character to the crystal. The anion layers (Sb-square net and CrSb$_2$-layers) are separated by a cation La layer. These atomic arrangements are responsible for anisotropic electrical and magnetic properties.[24] From Figure 1c (left panel), electronic charge density is localized on Cr and La atoms. The 3$d^3$ states of Cr$^{3+}$ in each CrSb$_6$ octahedron experience the highest crystal field energy, and they favorably split into $e_g$ and $t_{2g}$ energy levels, which are the main source of magnetism in this material (Figure 1c right panel). These levels are split further among themselves. The orbitals $d_{x^2-y^2}$ and $d_{z^2}$ split about 1.0 eV above and below the $E_F$, respectively and give rise to 1 $\mu_B$ of magnetism. However, the $t_{2g}$ states split and are localized in the order of meV around the $E_F$. This adds a small magnetic moment to the Cr atoms, due to the $E_F$ lying in the middle of the band. Therefore, LaCrSb$_3$ is known to exhibit itinerant ferromagnetism.[25] The transition temperature $T_C$ is 125 K, and their spins are aligned in the $b$–$c$ plane.[26] This means that the $b-$ and $c-$axes are the easy axes, whereas the $a-$axis is the hard axis. Below 95 K, spins point 18° away from the $b-$axis in the $b$–$c$ plane, reminiscent of an antiferromagnetic (AFM) component along the $c-$axis. Figure S2 shows the temperature-dependent resistivity behavior along different crystallographic axes; the resistivity rapidly decreases after $T_C$. Magnetization measurements are consistent with the $b$-axis being the easy axis; $T_C$ also corresponds to that observed in the resistivity measurement (Figure S9).

Electrical resistivity is directly related to the density of states of materials at $E_F$, whereas AHE is controlled by the BC concerning all the occupied states below $E_F$. We measured the Hall resistivity $\rho_H$ and longitudinal resistivity $\rho$ as a function of field $B$ along all three crystallographic axes of LaCrSb$_3$ at varying temperatures, as shown in **Figure 2**a and Figure S3, respectively. $\rho_H$ shows anomalous behavior up to $T_C$ (~125 K) along the $b-$ and $c-$axes (Fig. S6); their respective anomalous values at 2 K are 1.2 and 0.32 µΩcm, whereas the $a-$axis data do not show any anomalous behavior (Figure 2a). The Hall conductivity is calculated according to the relation: $\sigma_{xy} = \frac{\rho_{yx}}{\rho_{yx}^2 + \rho_{xx}^2}$. In Figure 2b, the corresponding anomalous Hall conductivity (AHC) are 1250 Ω$^{-1}$cm$^{-1}$ for the $b-$axis and 1150 Ω$^{-1}$cm$^{-1}$ for the $c-$axis at 2 K; which gradually decrease to zero as the temperature approaches $T_c$ (Figure 2c). It should be noted that the Hall conductivity remains zero up to 0.17 T along $c-$axis after which it suddenly rises to attain the saturation. This behavior is consistent with the small AFM interaction along $c-$axis owing to the fact that the anomalous velocity over all the occupied states in an AFM is zero. It is clear that the measured AHE is strongly anisotropic and appears only when $B$ applies in the $b$–$c$ plane. Magnetic field-dependent magnetization measurements for LaCrSb$_3$ at 2 K along different crystallographic axes are shown in Figure 2d. The magnetic moments are easily aligned along the $b-$ and $c-$axes, whereas $a-$axis is the hard axis, evidencing an anisotropic magnetic behavior. As compared to $b-$axis, magnetization along $c-$axis starts to increase slowly at small field, then suddenly jumps to saturation, accounting for the 18° spin canting towards the $c-$axis in the $b$–$c$ plane. The saturation magnetization reaches 1.6 µ$_B$/f.u. for LaCrSb$_3$, which is the same as previously reported.[25, 26]

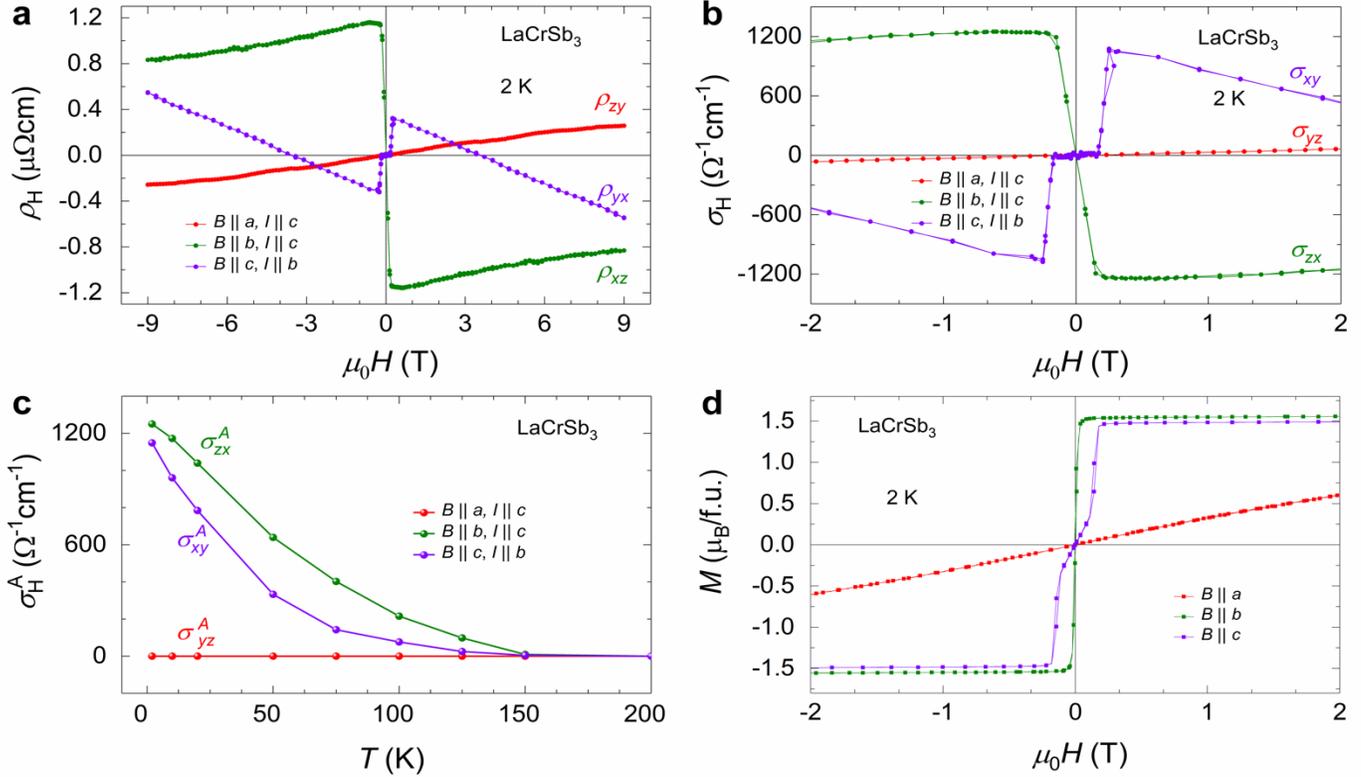

**Figure 2.** Hall resistivity $\rho_H$, Hall conductivity $\sigma_H$, and magnetization M of LaCrSb$_3$. a) Field-dependent behavior of $\rho_H$ along the $a-$, $b-$ and $c-$axes at 2 K. The anomalous value of $\rho_H$, i.e., $\rho_H^A$, can be derived by extrapolating the high-field part of $\rho_H$ to zero field. The field along the $a$–axis does not show anomalous behavior in $\rho_H$. b) The corresponding estimated value $\sigma_H$ to measured data in a. The anomalous value of $\sigma_H$, i.e., $\sigma_H^A$, can also be estimated similar to $\rho_H^A$. c) $\sigma_H^A$ appears in both directions ($b-$ and $c-$axes) up to 125 K, below the magnetic transition of LaCrSb$_3$. For example, $\sigma_{xy}$ is calculated from the relation $\sigma_{xy} = \frac{\rho_{yx}}{\rho_{xx}^2+\rho_{xy}^2}$. d) Field-dependent magnetic measurements at 2 K along three crystallographic axes. The magnetic moments are easily aligned along the $b-$ and $c-$axes and exhibit a similar saturation magnetization of 1.6 $\mu_B$/f.u., whereas the $a-$axis is a hard axis.

To gain an insight into the giant observed AHC, we used constrained-moment *ab-initio* calculations with the local spin density approximation (LSDA) exchange-correlation potential to simulate the band structure of LaCrSb$_3$ with the experimental lattice parameters and magnetic moments with spins pointing along $b$-axis. We calculated the momentum space BC of the electronic structure, revealing a large volume, which is centered on the $\Gamma$ point (**Figure 3**a). This originates further due to the inversion between Cr-$d$ and Sb-$p$ orbitals, forming rather dispersionless bands around this region. We found two interesting features: i) Nontrivial bands in the plane $k_y$-$k_z$ that produce large non-zero BC.[10, 11] This large volume of BC is different from normal magnetic metals, which show delta-like "hot" spots in the Brillion zone (BZ), for example, *bcc* Fe.[2] Such a unique and large volume of distribution provides a giant AHC. ii) Trivial bands along $k_x$ due to the weak coupling between Cr-Sb layers and La layers, producing the large longitudinal resistivity and negligible anomalous Hall effect as we have observed. Figures 3b-d show the electronic band structures for LaCrSb$_3$ along the high symmetric points (shown by dashed green, black and pink lines), depicting many nontrivial nearly dispersionless bands in the vicinity of $E_F$. From their atomic orbital contributions as given in Figure S14, these bands are mainly Cr-$d$ dominated, which play a crucial role for the electric and magnetic properties. These nontrivial bands are localized around the center of the Brillouin zone along $k_x$. A recent study [27] on the effect of reducing band width in the enhancement of AHC is consistent with our observation of large AHC in quasi-dispersionless bands in LaCrSb$_3$

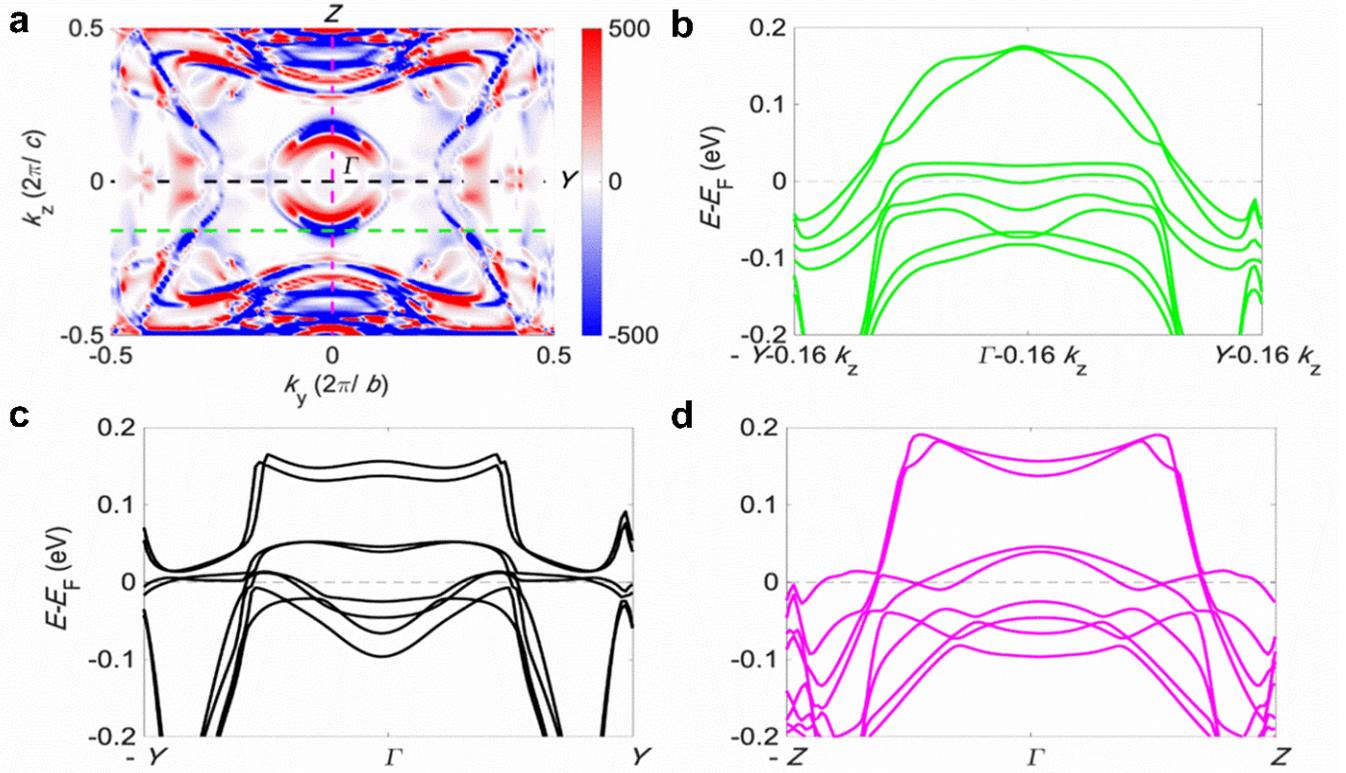

**Figure 3.** Berry curvature (BC) and corresponding bands of LaCrSb$_3$. a) Showing the large value of BC distribution of $k_y$-$k_z$ plane red as positive BC and blue as negative, in which plane the giant anomalous Hall conductivity is observed. The band energy dispersion corresponding the color dashed lines along b) Z-S (a path shifted -0.16$k_z$ from $\Gamma$-Y), c) $\Gamma$-Y, d) $\Gamma$-Z high symmetry points of Brillouin zone, indicating the large value of BC is exhibited by the nontrivial less-dispersive bands.

(see Figure S15 for a tight binding method based toy model).

**Effect of *f*-electron**: After measuring the remarkable values of AHC in the parent compound LaCrSb$_3$, it is highly desirable to obtain understanding about the AHE behavior by measuring other compounds from $R$CrSb$_3$ series, where $R$ is a rare earth element. Therefore, we extended our study to CeCrSb$_3$ and NdCrSb$_3$ possessing *f*-electrons and measured the field-dependent $\rho_H$ and $\rho$ values (Figure S4–S7) at various temperatures. For CeCrSb$_3$, the measured anomalous values of $\rho_H$ along *b*– and *c*–axes at 2 K are 4.3 µΩ cm and 5.2 µΩ cm, respectively (see Figure S6). The resulting values of $\sigma_H$ are plotted in **Figure 4** at various temperatures, which shows that the anomalous behavior of $\sigma_H$ is quite similar to that of LaCrSb$_3$. The AHC of CeCrSb$_3$ at 2 K is 1550 Ω$^{-1}$cm$^{-1}$ along both the *b*– and *c*–axes (Figure 4a), and the AHC of NdCrSb$_3$ at the same temperature is 2900 Ω$^{-1}$ cm$^{-1}$ for the *b*–axis (Figure 4b) and 900 Ω$^{-1}$cm$^{-1}$ for the *c*–axis. As expected, these values decrease as temperature reaches at $T_C$ (Figure 4c). Even though *f*-electrons introduce finite spontaneous magnetization along *a*–axis in CeCrSb$_3$ and NdCrSb$_3$, AHE is negligible for both compounds over the entire temperature range, as observed with LaCrSb$_3$. This demonstrates that Cr–*d* electrons are largely responsible for the AHE in this series of compounds. Moreover, there is no one-to-one correspondence between observed AHC and magnetic moment along various axes. This can be seen from the column plot in Figure 4d, signifying the role of electronic structure for AHC. For example, even though the magnetization of NdCrSb$_3$ is the highest along *a*–axis at 2K but produces negligible AHC. The similar effect is also observed for CeCrSb$_3$. Among the series of compounds, NdCrSb$_3$ shows giant AHC, which is the highest ever measured in any material to the best of our knowledge. NdCrSb$_3$ has a larger moment as compared to LaCrSb$_3$ and CeCrSb$_3$, and that moment is weakly coupled to the Cr $d-$states. This is evident from a metamagnetic transition in magnetization and AHE data of NdCrSb$_3$ along $b-$axis. However, above the ordering temperature of Nd spins (12 K), the effect is lost and the behavior and value of AHC closely follow that of LaCrSb$_3$. Larger magnetic moment in NdCrSb$_3$ compared to LaCrSb$_3$ and CeCrSb$_3$ at low temperature could be one of the reasons for the existence of giant AHC. The observation that the AHC in NdCrSb$_3$ decreases sharply and closely follows that of LaCrSb$_3$ after Nd-spin ordering temperature, indicates that Cr-d electrons dominated in the larger part of the temperature range studied. Owing to the correlation effect for added $f$-electrons in CeCrSb$_3$ and NdCrSb$_3$, it is not straightforward to estimate AHC from first principles calculations. It should be noted that the sign of anomalous Hall resistivity differs for various axes in all three compounds despite maintaining the same measurement geometry. Hence, this sign is dictated by the sign of Berry curvature for a particular direction of applied magnetic field. For a broad comparison, the observed values of AHC of some notable compounds are shown in Figure S13.

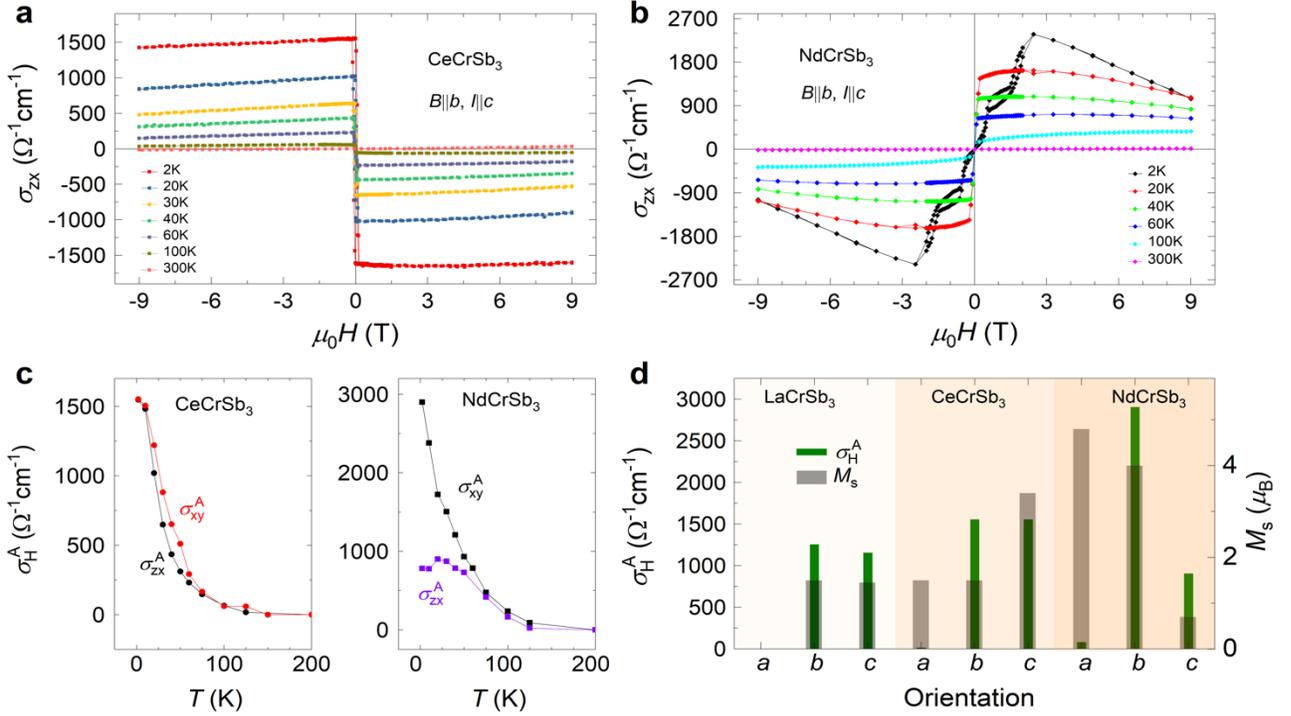

***Figure 4.*** *Hall conductivity $\sigma_{zx}$, anomalous Hall conductivity $\sigma_{zx}^A$ of CeCrSb$_3$ and NdCrSb$_3$ and their $\sigma_H^A$ value with saturation magnetization $M_s$. a) $\sigma_H$ along $B \parallel b$, $I \parallel c$ for CeCrSb$_3$. b) $\sigma_H$ along $B \parallel b$, $I \parallel c$ for NdCrSb$_3$. These measurements clearly show that the anomalous value of $\sigma_H$ smoothly changes across the transition temperature of rare-earth metals, reflecting the negligible effect of their spins. c) Temperature-dependent evolution of $\sigma_H^A$ of CeCrSb$_3$ (left panel) and NdCrSb$_3$ (right panel) along the $b-$ and $c-$axes up to 200 K. The non-zero values of $\sigma_H^A$ appear only below their magnetic transitions. d) Column plot of measured $\sigma_H^A$ and $M_s$ at 2 K with different orientations for LaCrSb$_3$, CeCrSb$_3$ and NdCrSb$_3$. Among them, NdCrSb$_3$ shows the largest value of $\sigma_H^A$.*

The anomalous behaviors in $R$CrSb$_3$ not only appear in electrical transport but are also found in thermal transport, i.e., the Nernst thermopower $S$. Figure S8 shows the field-dependent measured $S_{xz}$ of CeCrSb$_3$ at different temperatures, affirming an anomalous behavior. The anomalous value of $S_{xz}$ is found to be 2.5 µVK$^{-1}$ at 21 K, which associates these compounds with the non-trivial materials, exhibiting high anomalous Nernst effect.[20, 28]

Another important parameter, anomalous Hall angle (AHA), defines as how much longitudinal current coverts into the transverse direction. The estimated AHA is 4–10% for the present series of compounds (see **Table 1**). It is notable that despite a large carrier concentration $\sim 10^{22}$ cm$^{-3}$, the value of AHA compares to materials like

Weyl semimetal $Co_3Sn_2S_2$ where the carrier concentration is at least two orders of magnitude smaller.[21, 29]

**Discussion**
$R$CrSb$_3$ is a promising series of quasi-2D compounds that exhibit high anisotropic AHC. The measured AHC is the sum of all the contributions from the entire BZ and that can have both intrinsic and extrinsic origins. From the framework of unified models that are valid for the varieties of compounds having conductivity beyond the range of $10^4 < \sigma < 10^6$ $\Omega^{-1}$cm$^{-1}$, extrinsic origins dominate.[30, 31] The conductivity of the $R$CrSb$_3$ series of compounds ranges from $0.7 \times 10^4$ to $5.9 \times 10^5$ $\Omega^{-1}$cm$^{-1}$, which lie within the moderate range of conductivity. The temperature-dependent data of $\sigma_H^A$ vs $\sigma$ is neither constant nor linear, excluding the single contribution from the Berry phase or skew scattering (Figure S12). However, from the power law behavior $\sigma_H^A \propto \sigma^n$, $n$ is found to be 1.7 for LaCrSb$_3$ and CeCeSb$_3$ (Figure S12). For cases where a mixed contribution of Berry phase and side-jump dominates, $n$ is predicted to be 1.6.[32] Surprisingly, our estimation of $n = 3$ for NdCrSb$_3$ goes beyond this power law and calls for more accurate scaling law. Due to nonlinear behavior of $\sigma_H^A$ vs $\sigma^2$, it is hard to estimate the accurate value of AHC from Berry phase by a linear intercept method. To naively check, a rough linear intercept of $\sigma_H^A$ vs $\sigma^2$ for low temperature region gives rise to 220 $\Omega^{-1}$cm$^{-1}$ intrinsic AHC for LaCrSb$_3$. However, if one considers the case of ultraclean limit for the present systems, the predicted AHC from skew scattering is 630 $\Omega^{-1}$ cm$^{-1}$ (from the relation $e^2/ha$ for $a = 6.16$ Å in LaCrSb$_3$) whereas the side-jump contributions are much smaller than $e^2/ha$, i.e., by an order of $\frac{\varepsilon_{SO}}{E_F} \sim 10^{-1} - 10^{-3}$. Hence, even if one applies the extreme case of ultraclean limit in LaCrSb$_3$, the maximum extrinsic AHC contribution empirically is ~ 700 $\Omega^{-1}$ cm$^{-1}$. This still amounts to a minimum intrinsic AHC of ~550 $\Omega^{-1}$ cm$^{-1}$. It indicates that the linear intercept of $\sigma_H^A$ vs $\sigma^2$ is not a good approximation for estimating intrinsic contribution of AHC in this system. Furthermore, AHE derived from BC is resonantly enhanced when $\frac{\hbar}{\tau}$ and $\varepsilon_{SO}$ of materials are equivalent, i.e., $\frac{\hbar}{\tau} \cong \varepsilon_{SO}$, where $\tau$ is relaxation time, $\hbar$ is the reduced Planck constant, and $\varepsilon_{SO}$ is SOC energy of bands close to $E_F$.[30] We found that the values of $\frac{\hbar}{\tau}$ and $\varepsilon_{SO}$ are 0.7 and 0.8, respectively, for the $R$CrSb$_3$ series of compounds, and they are best matched in the criteria of the resonantly enhanced AHE. These scenarios indicate that the measured AHC of $R$CrSb$_3$ can arise from a mixture of intrinsic and extrinsic origins that is hard to separate out. The low temperature longitudinal conductivity of all the three compounds is between $10^4$ to $10^5$ $\Omega^{-1}$cm$^{-1}$ which is still in the moderate conductivity limit, but is close to the boundary of dirty limit. Hence, the most probable cause of the extrinsic contribution can be the side-jump effect. It originates from the change in the momentum of the Gaussian wave packet when it interacts to a sufficiently smooth impurity potential in the presence of spin-orbit interaction.[3] Like the intrinsic effect, it is also independent of the scattering time and hence very difficult to differentiate.

**Conclusion**
We observed large values of anisotropic AHC in $R$CrSb$_3$ ($R$ = La, Ce and Nd) series of compounds. Effect of the introducing $f$-electrons as in CeCrSb$_3$ and NdCrSb$_3$ shows enhancement in AHC. The large magnetic moment in NdCrSb$_3$ can be one of the reasons for the existence of giant AHC. We demonstrate that power law scaling for anomalous Hall conductivity follows $\sigma_H^A \propto \sigma^{1.7}$ which is valid in the intrinsic and side-jump regime for LaCrSb$_3$ and CeCrSb$_3$ while it goes beyond this scaling for NdCrSb$_3$. The positive aspects of the existence of rather dispersionless bands for observing large value of anomalous Hall conductivity have also been discussed for the first time which will provide motivation for exploring anomalous transport in flat-band magnetic systems.

*Table 1.* *Anomalous values of studied compounds at 2K. Anomalous Hall resistivity (AHR), anomalous Hall conductivity (AHC), resistivity (ρ), charge carrier density (n), anomalous Hall angle (AHA), ferromagnetic transition temperature ($T_c$).*

| Compound | B ∥ | $T_c$ (K) | ρ ($10^{-5} \times \Omega$cm) | n ($10^{22} \times$ cm$^{-3}$) | AHR (μΩcm) | AHC ($\Omega^{-1}$cm$^{-1}$) | AHA (%) |
|---|---|---|---|---|---|---|---|
| LaCrSb$_3$ | *a*-axis |  | 30.27 |  | 0 | 0 | 0 |
|  | *b*-axis | 130 | 1.75 | 2.12 | 1.21 | 1250 | 4 |
|  | *c*-axis |  | 3.04 |  | 0.32 | 1150 | 1.8 |
| CeCrSb$_3$ | *a*-axis |  |  |  | 0.19 | 10 | 0.18 |
|  | *b*-axis | 125 | 5.33 | 0.92 | 4.3 | 1550 | 8.1 |
|  | *c*-axis |  | 5.19 |  | 5.2 | 1550 | 10.0 |
| NdCrSb$_3$ | *a*-axis |  |  |  | 0.8 | 75 | 0.77 |
|  | *b*-axis | 116 | 3.26 | 1.62 | 2.5 | 2900 | 7.7 |
|  | *c*-axis |  | 4.98 |  | 1.2 | 900 | 2.5 |

**Acknowledgements**

This work was financially supported by the European Research Council (ERC) Advanced Grant No. 742068 ("TOPMAT") and Deutsche Forschungsgemeinschaft DFG under SFB 1143 (project no. 247310070).

**Conflict of Interest**

The authors declare no conflict of interest.

# Supplementary Information

**Materials synthesis and characterizations**

Single crystals of $R$CrSb$_3$ (where $R$=La, Ce, Nd) were grown by flux-growth method using Sb as a flux. For LaCrSb$_3$, La polished pieces (~ 99.9 %), Cr pieces (~ 99.99 %) with Sb shots (~ 99.999 %) were initially loaded in an alumina crucible in a molar ratio of 1:1:10 and sealed in an evacuated quartz tube under 3-mbar partial pressure of argon. The ampoule was heated at a rate of 100 K/h up to 1372 K and kept for 12 h at this temperature. For the single crystal growth, the melt was slowly cooled down to 972 K at a rate of 2 K/h and at this temperature, the extra Sb flux was removed with the help of centrifuge. Same temperature profile for CeCrSb$_3$ and NdCrSb$_3$ single crystal was used with a slightly different molar ratio i.e. Ce:Cr:Sb::1:1:12, Nd:Cr:Sb::1:2:20. From this method, typical 4-8 mm large crystals were obtained, which were used for further measurements.

The detailed compositional and structural analyses were investigated by energy-dispersive X-ray analysis and single crystal X-ray diffraction, respectively. The preferred growth orientation of $R$CrSb$_3$ crystals was [100] as confirmed by Laue diffraction. For further transport measurements, crystals were orientated with the help of Laue diffraction and were cut into rectangular bars in desired direction using the wire saw.

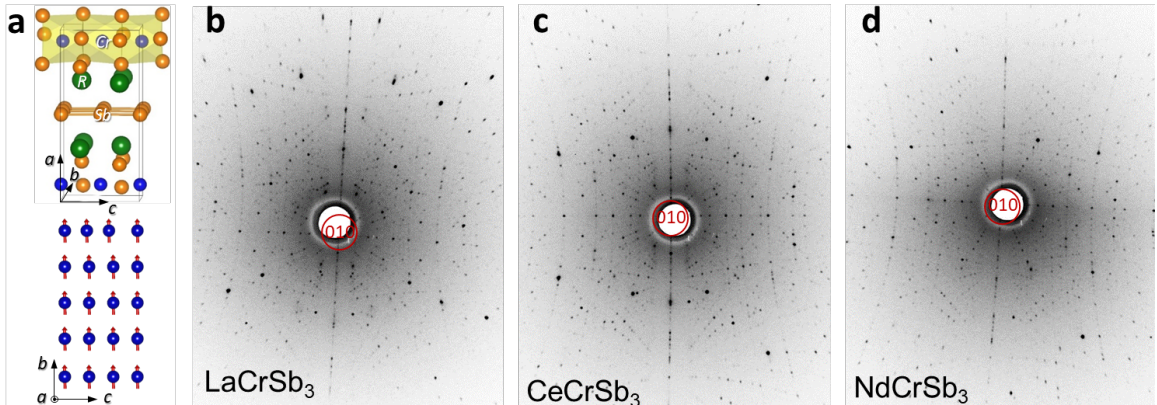

*Figure S1.* Crystal and magnetic structures and Laue diffraction of $R$CrSb$_3$. a) (top) Unit cell of $R$CrSb$_3$, indicating rare-earth R (green), Cr (blue), and Sb (orange). CrSb$_6$ octahedra along the b– and c–axes share an edge and face, respectively, which are responsible for crystal and magnetic anisotropy. (bottom) Magnetic structure of $R$CrSb$_3$ pointing Cr spins along the b–axis (for LaCrSb$_3$, these spins are 20° off from the b–axis). Here, spin of R is not considered. Laue diffraction patterns focusing X-ray along b-axis b) for LaCrSb$_3$, c) CeCrSb$_3$, and d) NdCrSb$_3$.

**Electrical transport measurements**

**(A) Zero-field resistivity and carrier density**

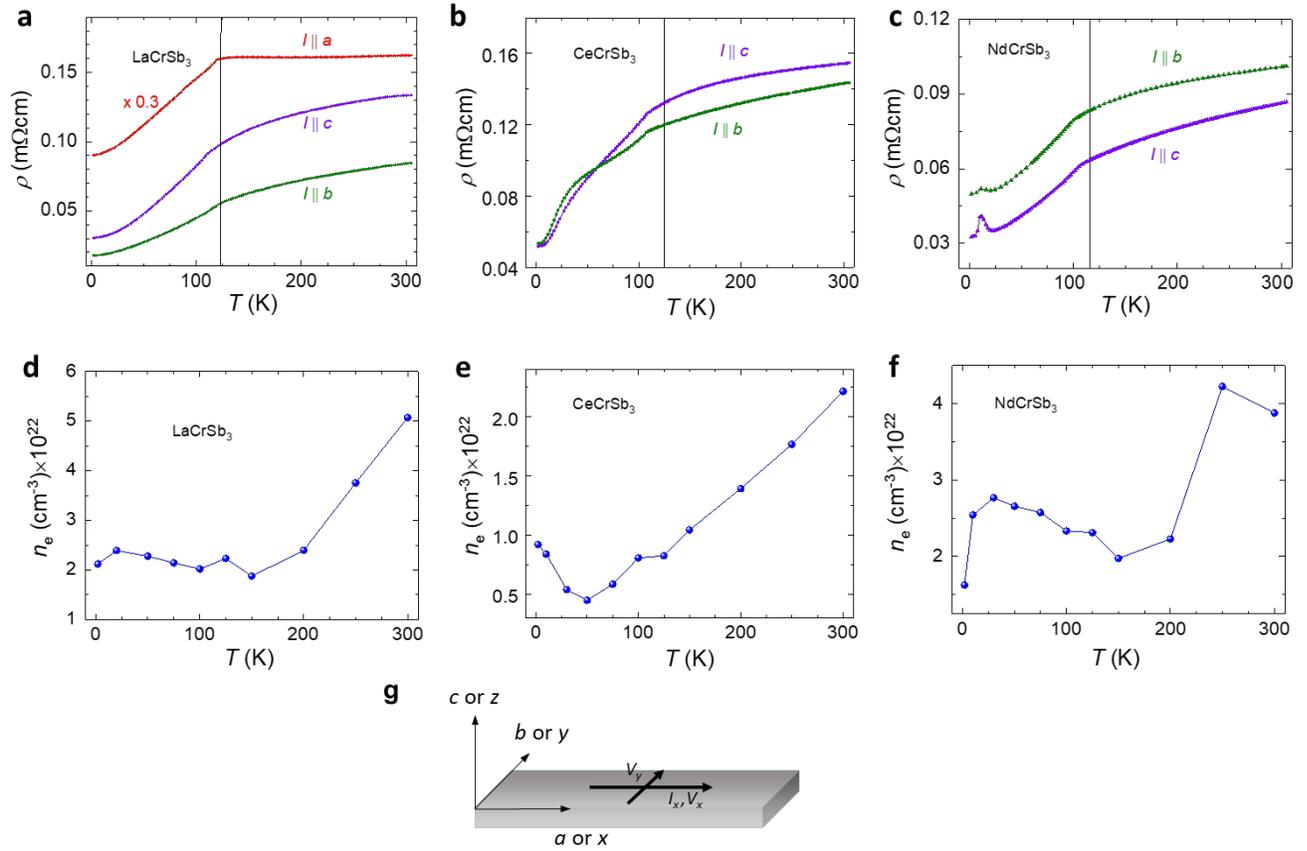

*Figure S2* Zero-field longitudinal resistivity ($\rho_{xx}$), carrier concentration and measurement geometry. a), b), c) Temperature dependent zero field resistivity along different crystallographic axes and d), e), f) carrier concentration ($n_e$) of LaCrSb$_3$, CeCrSb$_3$ and NdCrSb$_3$, respectively. g) Sketch representing a-, b- and c-axes in the equivalent Cartesian coordinate system.

## (B) Field dependent resistivity

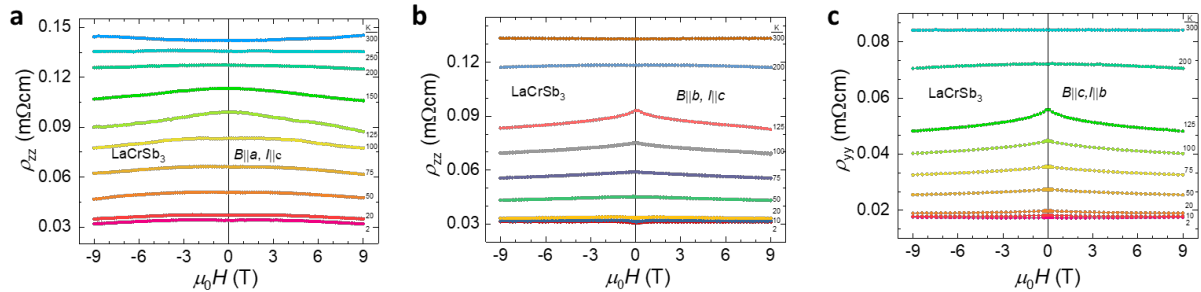

*Figure S3.* Magnetic field dependence of resistivity with field along different crystallographic a−, b− and c−axes at various temperatures for LaCrSb$_3$.

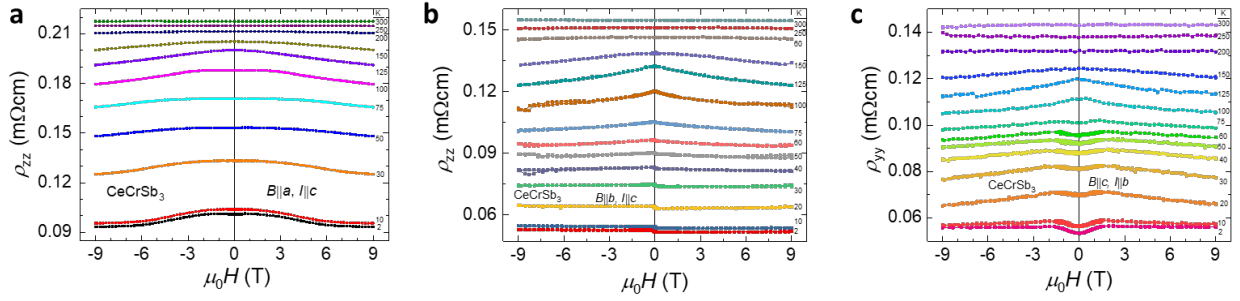

*Figure S4.* Magnetic field dependence of resistivity with field along different crystallographic a−, b− and c−axes at various temperatures for CeCrSb$_3$.

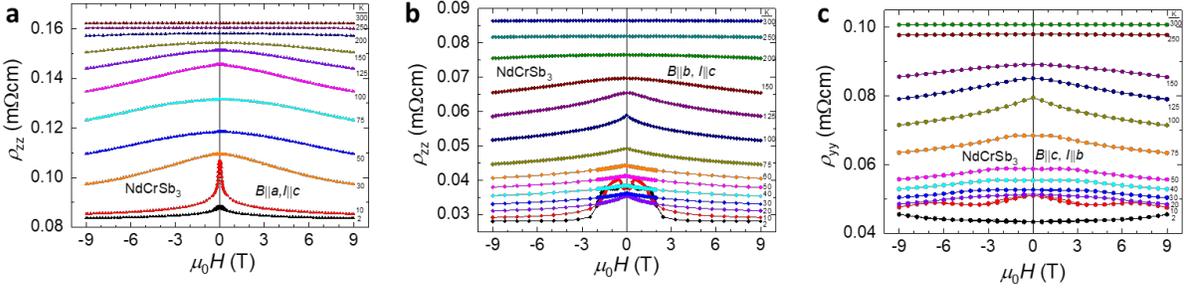

*Figure S5.* Magnetic field dependence of resistivity with field along a) a−, b) b−, c) c−axes for NdCrSb$_3$ at different temperatures.

## Hall resistivity

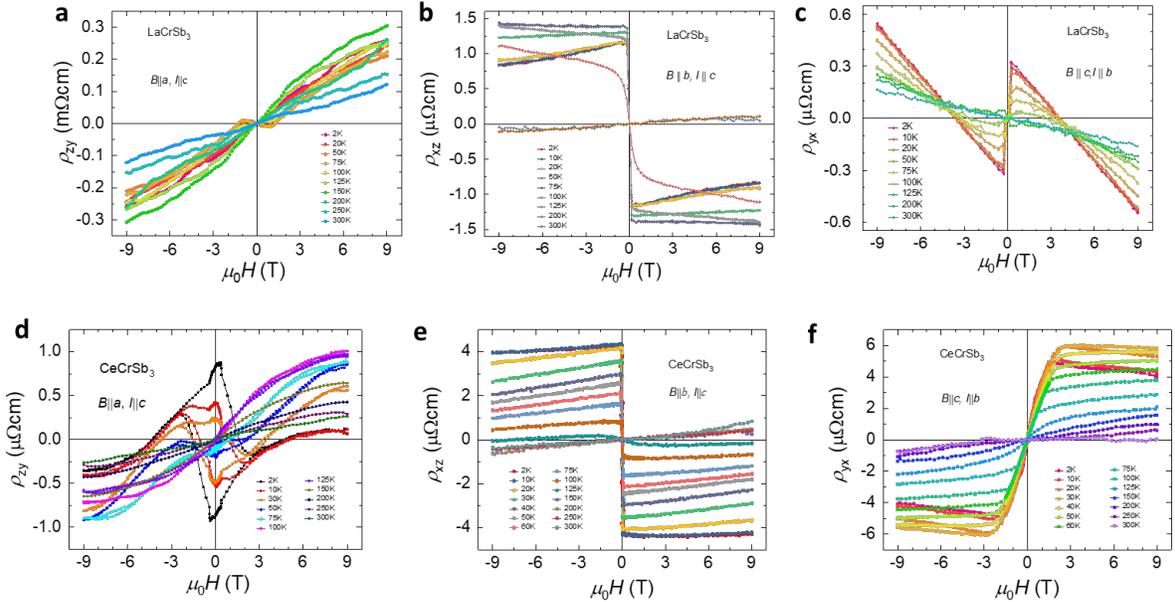

*Figure S6.* Magnetic field dependent Hall resistivity for the field along a) a−, b) b−, c) c−axes for LaCrSb$_3$ and d) a−, e) b−, f) c−axes for CeCrSb$_3$ at different temperatures.

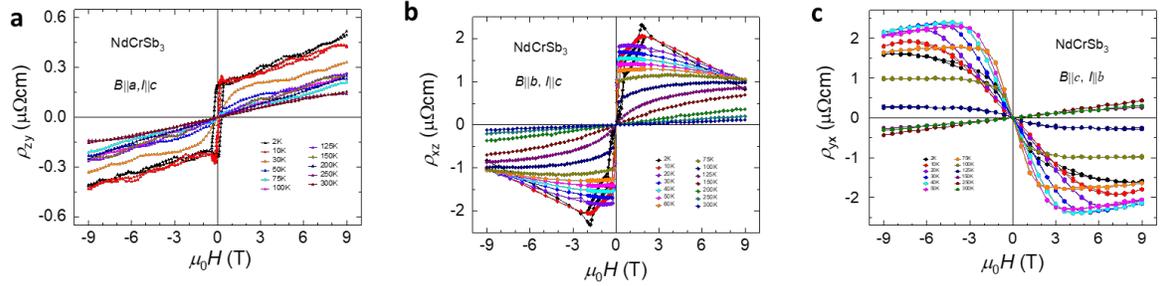

*Figure S7.* Magnetic field dependent behavior of Hall Resistivity for the field along a) a−, b) b−, and c) c−axes at different temperatures for NdCrSb$_3$.

## Thermal transport measurements

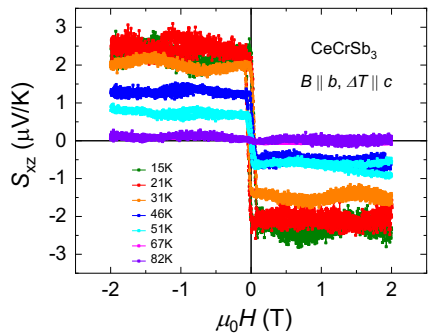

*Figure S8.* Nernst thermopower ($S_{xz}$) for CeCrSb$_3$. Magnetic field dependence of $S_{xz}$ for CeCrSb$_3$. The magnetic field is applied along b−axis and temperature gradient along c-axis.

## Magnetic measurements

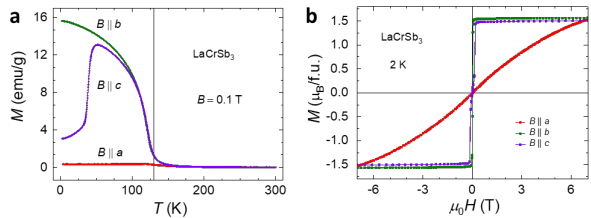

*Figure S9.* Magnetization of LaCrSb$_3$. a) Temperature dependence of magnetization, and b) magnetic field dependence of magnetization for LaCrSb$_3$ along three different crystallographic axes.

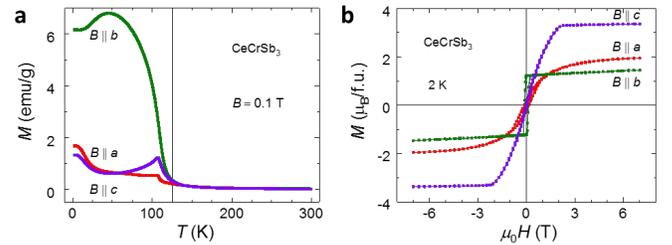

*Figure S10.* Magnetization of CeCrSb$_3$. a) Temperature dependence of magnetization, and b) magnetic field dependence of magnetization for CeCrSb$_3$ along three different crystallographic axes.

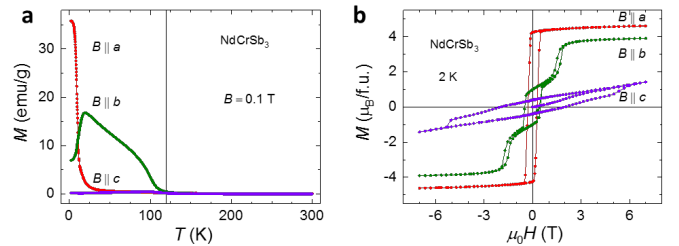

*Figure S11.* Magnetization of NdCrSb$_3$. a) Temperature dependence of magnetization, and b) magnetic field dependence of magnetization for NdCrSb$_3$ along three different crystallographic axes.

## Power scaling law

The recent numerical calculations of $\sigma_{xy}^A$ for various regimes of $\sigma_{xx}$ have been done by Onoda et al. (1, 2) Depending on the range of $\sigma_{xx}$, $\sigma_{xy}^A$ varies with different power of $\sigma_{xx}$ and covers a broad range of sctatterings i.e. $\sigma_{xy}^A \propto \sigma_{xx}^n$. For the spesific value of *n*, the followings regimes of $\sigma_{xx}$ are defined in the literatures.

1) A high conducting regime $\sigma_{xx} > 10^6$ $\Omega^{-1}$cm$^{-1}$ in which a linear contribution to $\sigma_{xy}^A$ due to skew scattering (extrinsic $\sigma_{xy}^A$) dominates i.e. $\sigma_{xy}^A \propto \sigma_{xx}$. This directly indicates that anomalous Hall angle ($\sigma_{xy}^A/\sigma_{xx}$) is constant throughout the variation.
2) A moderate conducting regime $10^4 < \sigma_{xx} < 10^6$ $\Omega^{-1}$cm$^{-1}$, where scattering-independent intrinsic contribution to $\sigma_{xy}^A$ dominates i.e. $\sigma_{xy}^A$ = constant. However, such regime has never been observed in broader range in a real material.
3) A bad metallic regime $\sigma_{xx} < 10^4$ $\Omega^{-1}$cm$^{-1}$, where side-jump (extrinsic $\sigma_{xy}^A$) dominates and $\sigma_{xy}^A$ depends on $\sigma_{xx}$ with a power of 1.6 i.e. $\sigma_{xy}^A \propto \sigma_{xx}^{1.6}$.

In the above criteria, $\sigma_{xx}$ is considered to be sample-dependent (residual value). However, for a fixed sample, $\sigma_{xx}$ will vary when temperature dependent $\sigma_{xy}^A$ is measured. The energy broadening of bands which is related to scattering time, increases with increasing temperature and when it is larger than the energy splitting of bands due to spin-orbit interaction, $n$ reaches to 2 i.e. $\sigma_{xy}^A \propto \sigma_{xx}^2$ as revealed by Berry phase induced intrinsic behavior (3). Interestingly, the value of $n$ is found to be $n = 1.7$ for LaCrSb$_3$, $n = 1.7$ for CeCrSb$_3$ and $n = 3$ for NdCrSb$_3$, which are below and above the Berry phase induced AHC. The contributions from Berry phase AHC can be deduced from the intercept of straight line of $\sigma_{xy}^A$ vs $\sigma_{xx}^2$ plot. However, due to a nonlinear behavior in the whole temperature range of measurement, this intercept method is not justified in the present case. While doing so, the intercept from the low temperature data reveals the value of AHC 220 $\Omega^{-1}$ cm$^{-1}$ for LaCrSb$_3$ Fig. S12.

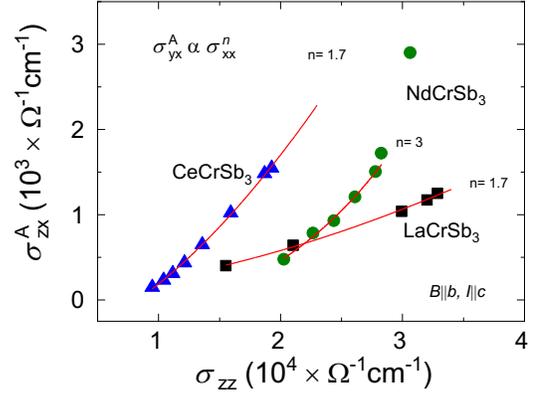

**Figure S12.** *A power law scaling between the anomalous Hall resistivity ($\sigma_{xz}^A$) to resistivity ($\sigma_{zz}$) for LaCrSb$_3$, CeCrSb$_3$ and NdCrSb$_3$. Each data point belongs to different temperatures below $T_c$.*

**Anomalous Hall value for selected compounds**

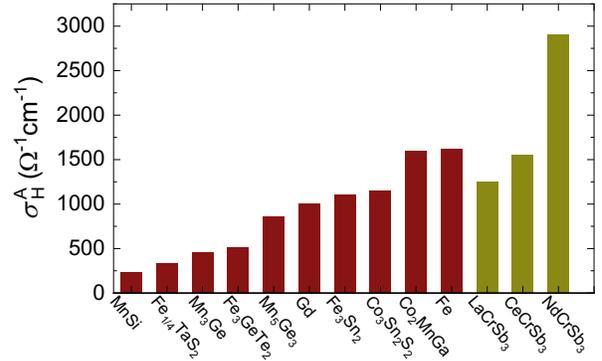

**Figure S13.** *Column plot of $\sigma_H^A$ for some selected compounds and elements that exhibit anomalous Hall conductivity; the column data were taken from Table S1. Among them, NdCrSb$_3$ shows the largest value of $\sigma_H^A$.*

**Band structure of LaCrSb$_3$**

Bands around Fermi level, $E_F$, are dominated by the Cr-$d$ orbitals, with $d_{yz}$ crossing $E_F$, and a slight hybridization with the Sb-$p_y$ orbital. The La-$f$ states are highly localized around 1.5 to 2.0 eV above the Fermi level. Although the La-$f$ states are not occupied, they play a crucial role with that of the Sb-$p$ to sandwich the Cr-$d$ states to be more localized around the $E_F$. Since the f-states above $E_F$ is almost forbidden for the Cr-d electrons, the energy window for Cr-d states is compressed and results in the Cr-d dominated flat bands around $E_F$. Furthermore, because of the orthorhombic crystal structure, the crystal field causes the $e_g$, and $t_{2g}$ states to split. These $e_g$, and $t_{2g}$ states are further split among themselves, where the $d_{x^2-y^2}$ and $d_{z^2}$ are split on the order of 1 eV above and below the Fermi energy, respectively, to give 1 $\mu_B$ of magnetism. The $t_{2g}$ states are localized around the $E_F$ and split on the order of meV, with hybridization. This adds a small moment to the Cr atoms, due to the $E_F$ lying in the middle of the band.

## Effect of band flattening on anomalous Hall conductivity

The motivation to study the AHC in materials with quasi-dispersionless bands is based on the effective volume of band anti-crossings at the Fermi level. In most cases, the band anti-crossings are not only some isolated points but can also form some finite area in the reciprocal space. The overlap between band anti-crossing and Fermi level gives effective contributions to the intrinsic AHC. We would like to have such kind of overlaps as much as possible. Along with the dispersion of band structure, such kind of anti-crossing areas can also have dispersion in k-space. The overlap between band anti-crossings and Fermi level cannot be completely 100%. However, it is possible to increase the percentage of the overlap by reducing the band width with suitable chemical potential. This can be understood from an effective two band tight binding model

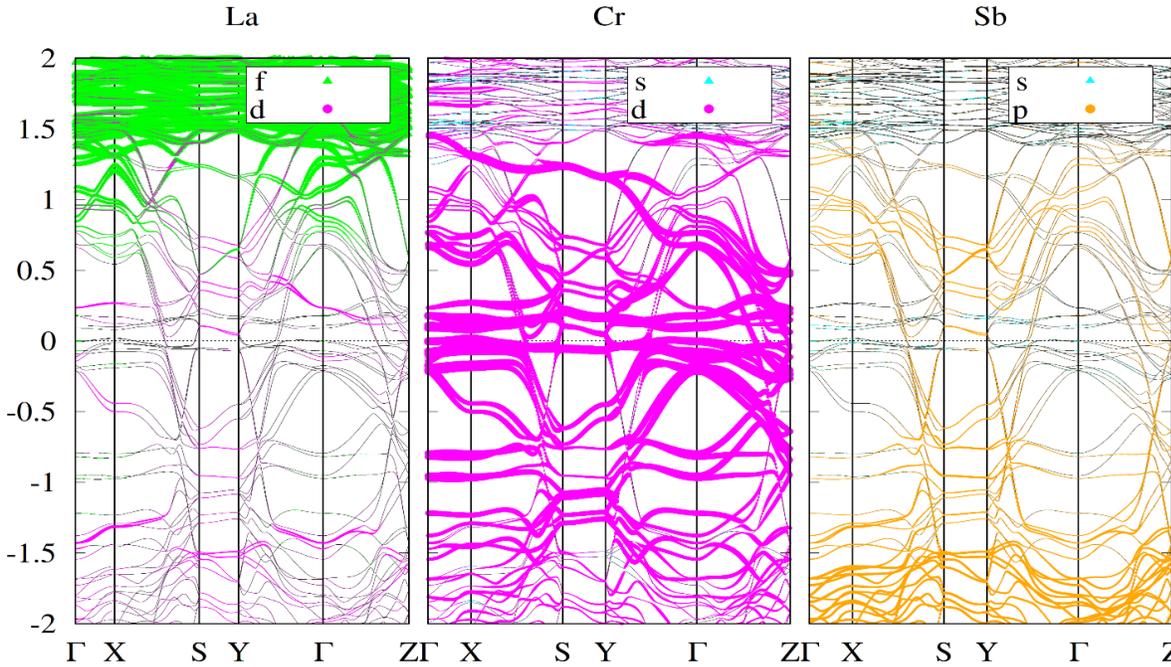

**Figure S14.** Atomic orbital and atom projected band structure of LaCrSb$_3$ along high symmetry lines.

$$H(\vec{k}) = \begin{pmatrix} M_1(k) & A_x \sin k_x + A_z \sin k_z - iA_y \\ A_x \sin k_x + A_z \sin k_z + iA_y \sin k_y & -M_2(k) \end{pmatrix}$$

, where

$$M_1(k) = M + B_x \cos k_x + B_y \cos k_y + B_z \cos k_z$$

,

$$M_2(k) = M' + B_x' \cos k_x + B_y' \cos k_y + B_z' \cos k_z$$

.

We have set the parameters to guarantee the existence of band anti-crossing between these two bands. Supposing only one electron in this system and keeping the existence of band anti-crossing, we decrease the band width of 1st band, see Figure S15 (a-f). For each case after tuning the band width, we have calculated the DOS and integrated it to get the chemical potential, such that always one electron is filled in the system. When the band-width of 1st band is large enough, the relationship between band-width and AHC is not so clear, such as from 2.7eV to 2.0 eV. However, as the band width becomes much smaller, the trend of AHC becomes much more obvious. From Figure S15 (g), one can find that the AHC increases dramatically with decreasing band width in the range of 2.0 eV to 0.2 eV. The effective change of AHC is 800 S/cm (from ~380 S/cm to ~1180 S/cm) by reducing the band width in the tight binding model. This accounts for an effective enhancement of more than 200 %. Indeed, this change is not small and is large enough to illustrate the important connection between the flatband and large intrinsic AHC.

This model analysis gives a clear understanding of our initial motivation.

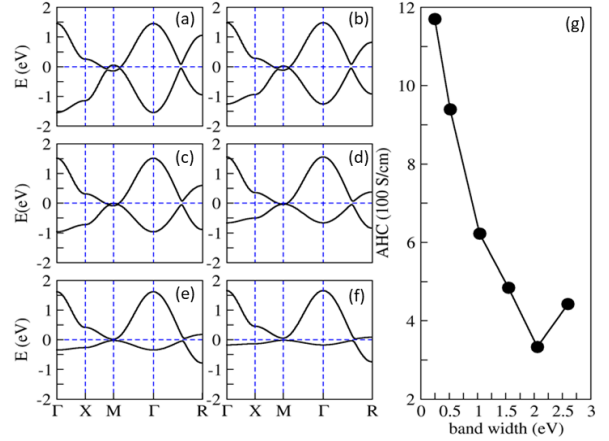

*Figure S15.* (a-f) Electronic band structure of the effective two band tight binding model. The parameters in (a) are $A_x$=0.01 eV, $A_y$=0.02 eV, $A_z$=0.015 eV, M=0.3 eV, $B_x$=0.1 eV, $B_y$=0.3 eV, $B_z$=0.25 eV, M'=0.2 eV, $B_x'$=0.3 eV, $B_y'$=0.1 eV, and $B_z'$=0.2 eV, respectively. To decrease the band width of the 1st band, the parameters of $B_x$, $B_y$, $B_z$ are (b) $B_x$=0.08 eV, $B_y$=0.24 eV, $B_z$=0.2 eV, (c) $B_x$=0.06 eV, $B_y$=0.18 eV, $B_z$=0.15 eV, (d) $B_x$=0.04 eV, $B_y$=0.12 eV, $B_z$=0.10 eV, (e) $B_x$=0.02 eV, $B_y$=0.06 eV, $B_z$=0.05 eV, and (f) $B_x$=0.01 eV, $B_y$=0.03 eV, $B_z$=0.025 eV, respectively. (g) AHC as a function of band width of 1st band.

*Table S1.* Topological nontrivial band derived anomalous Hall conductivity (AHC) in the compounds.

| Compound | \|AHC\| ($\Omega^{-1}$cm$^{-1}$) @2 K | Reference |
|---|---|---|
| LaCrSb$_3$ | 1150-1250 | Present work |
| CeCrSb$_3$ | 1550 | Present work |
| NdCrSb$_3$ | 900-2900 | Present work |
| MnSi | 230 | [4] |
| Fe$_{1/4}$TaS$_2$ | 336 | [5] |
| Mn$_3$Ge | 500 | [6] |
| Fe$_3$GeTe$_2$ | 540 | [7] |
| Mn$_5$Ge$_3$ | 860 | [8] |
| Gd-film | 1000 | [9] |
| Fe$_3$Sn$_2$ | 1100 | [10] |
| Co$_3$Sn$_2$S$_2$ | 1130 | [11] |
| Co$_2$MnGa | 1600, 2000 | [12, 13] |
| *bbc* Fe | 1030, 1670 (80 K) | [9, 14] |